\documentstyle[12pt]{article}
\font\twlmsb =msbm10 scaled \magstep1
\font\egtmsb =msbm8
\font\sevmsb =msbm7
\newfam\msbfam
\textfont\msbfam\twlmsb
\scriptfont\msbfam\egtmsb
\scriptscriptfont\msbfam\sevmsb
\def\Bbb{\protect\pBbb}
\def\pBbb{\relax\ifmmode\expandafter\Bb\else\typeout{You cann't use
Bbb in text mode}\fi}
\def\Bb #1{{\fam\msbfam\relax#1}}
\def\thebibliography#1{\bigskip\section*{\centering
References\\}\bigskip\list
{\arabic{enumi}.}{\settowidth\labelwidth{#1}\leftmargin\labelwidth
\advance\leftmargin\labelsep\usecounter{enumi}}
\def\newblock{\hskip .11em plus .33em minus .07em}
\sloppy\clubpenalty4000\widowpenalty4000 \sfcode`\.=1000\relax}

\def\op#1{\mathop{\fam0 #1}\limits}

\newcommand{\ben}{\begin{eqnarray}}
\newcommand{\een}{\end{eqnarray}}
\newcommand{\be}{\begin{eqnarray*}}
\newcommand{\ee}{\end{eqnarray*}}
\newcommand{\bea}{\begin{eqalph}}
\newcommand{\eea}{\end{eqalph}}
\newcommand{\cL}{{\cal L}}

\newcommand{\cH}{{\cal H}}
\newcommand{\cF}{{\cal F}}
\newcommand{\cD}{{\cal D}}
\newcommand{\al}{\alpha}
\newcommand{\bt}{\beta}
\newcommand{\dl}{\delta}
\newcommand{\la}{\lambda}

\newcommand{\om}{\omega}
\newcommand{\Om}{\Omega}
\newcommand{\m}{\mu}
\newcommand{\g}{\gamma}
\newcommand{\G}{\Gamma}

\newcommand{\si}{\sigma}
\newcommand{\Si}{\Sigma}

\newcommand{\w}{\wedge}

\newcommand{\wt}{\widetilde}
\newcommand{\wh}{\widehat}
\newcommand{\ol}{\overline}
\newcommand{\dr}{\partial}

\newcounter{eqalph}
\newcounter{equationa}

\newenvironment{eqalph}{\stepcounter{equation}
\setcounter{equationa}{\value{equation}}
\setcounter{equation}{0}

\begin{eqnarray}}{\end{eqnarray}
\setcounter{equation}{\value{equationa}}}

\begin{document}
\hbox{}

\centerline{\large\bf GAUGE GRAVITATION THEORY.}
\medskip

\centerline{\large\bf WHAT IS GEOMETRY OF THE WORLD?}
\bigskip

\centerline{\bf Gennadi A Sardanashvily}
\medskip

\centerline{Department of Theoretical Physics, Physics Faculty,}

\centerline{Moscow State University, 117234 Moscow, Russia}

\centerline{E-mail: sard@grav.phys.msu.su}
\bigskip

When joined the unified gauge picture of fundamental interactions, the
gravitation theory leads to geometry of a space-time which is far from
simplicity of pseudo-Riemannian geometry of Einstein's General Relativity.
This is geometry of the affine-metric composite dislocated manifolds.
The goal is modification of the familiar equations of a gravitational field
and entirely the new equations of its deviations. In the present brief,
we do not detail the mathematics, but discuss the reasons
why it is just this geometry. The major physical underlying reason lies in
spontaneous symmetry breaking when the fermion matter admits only the
Lorentz subgroup of world symmetries of the geometric arena.

\bigskip
{\bf I.}
\bigskip

Gauge theory is well-known to call into play the differential geometric
methods in order to describe interaction of fields possessing a certain
symmetry group. Moreover, it is this geometric approach phrased
in terms of fibred manifolds which provides
the adequate mathematical formulation of classical field theory.

Note that the conventional gauge principle of gauge invariance of
Lagrangian densities losts its validity since symmetries of realistic
field models are almost always broken. We follow the geometric
modification of the gauge principle which is formulated as follows.

Classical fields can be identified with sections of fibred manifolds
\begin{equation}
\pi: Y\to X \label{1}
\end{equation}
whose $n$-dimensional base $X$ is treated a parameter space,
in particular, a world
manifold. A locally trivial fibred manifold is called the fibre bundle
(or simply the bundle). We further provide $Y$ with an atlas of
fibred coordinates $(x^\mu, y^i)$ where $(x^\mu)$ are coordinates of
its base $X$.

Classical field theory meets the following tree types of fields as a rule:
\begin{itemize}
\item matter fields represented by sections of a vector bundle
\[
Y=(P\times V)/G
\]
associated with a certain principal bundle $P\to X$ whose structure group
is $G$;
\item gauge potentials identified to sections of the bundle
\begin{equation}
C:=J^1P/G\to X \label{3}
\end{equation}
of principal connections on $P$ (where $J^1P$ denotes the first order jet
manifold of the bundle $P\to X$);
\item classical Higgs fields described by global sections of the quotient
bundle
\[
\Si_K:= P/K\to X
\]
where $K$ is the exact symmetry closed subgroup of the Lie group $G$.
\end{itemize}

Dynamics of fields represented by sections of the fibred
manifold (\ref{1}) is phrased in terms of jet manifolds
\cite{gia,got,kol,kup,sard,lsar}.

Recall that the $k$-order jet manifold $J^kY$ of a fibred
manifold $Y\to X$ comprises the equivalence classes
$j^k_xs$, $x\in X$, of sections $s$ of $Y$ identified by the $(k+1)$
terms of their Taylor series at $x$. It is important that $J^kY$ is
a finite-dimensional manifold which meets all conditions usually
required of manifolds in field theory.
Jet manifolds have been widely used in the
theory of differential operators. Their application to differential
geometry is based on the  1:1 correspondence between the connections
on a fibred manifold $Y\to X$ and the global sections
\begin{equation}
\G=dx^\la\otimes(\dr_\la+\G^i_\la(y)\dr_i) \label{5}
\end{equation}
of the jet bundle
$J^1Y\to Y$ \cite{man,sard,sau}. The jet bundle $J^1Y\to Y$
is an affine bundle modelled on the
vector bundle
\[
\ol J^1Y=T^*X\op\otimes_Y VY
\]
where $VY$ denotes the vertical tangent bundle of the fibred manifold
$Y\to X$.
It follows that connections on a fibred manifold $Y$ constitute an
affine space modelled on the vector space of soldering forms
$Y\to \ol J^1Y$ on $Y$.

In the first order Lagrangian formalism,
the jet manifold $J^1Y$ plays the role of a finite-dimensional
configuration space of fields. Given fibred coordinates $(x^\la,y^i)$
of $Y\to X$, it is endowed with the adapted
coordinates $(x^\la,y^i,y^i_\la)$ where coordinates $y^i_\la$ make the
sense of values of first order partial derivatives $\dr_\la y^i(x)$ of field
functions $y^i(x)$.
In jet terms, a first order Lagrangian density of fields is represented
by an exterior horizontal density
\begin{equation}
L=\cL(x^\mu, y^i, y^i_\mu)\om, \qquad \om=dx^1\w...\w dx^n,  \label{6}
\end{equation}
on $J^1Y\to X$.

In field theory, all Lagrangian densities are polynomial forms
with respect to velocities $y^i_\la$.
Since the jet bundle $J^1Y\to Y$ is affine,
polynomial forms on $J^1Y$ are factorized by morphisms $J^1Y\to\ol J^1Y$.
It follows that every Lagrangian density of field theory is
represented by composition
\begin{equation}
L: J^1Y\op\to^D T^*X\op\otimes_Y VY\to\op\wedge^n T^*X \label{20}
\end{equation}
where $D$ is the covariant differential
\[
D=(y^i_\la-\G^i_\la)dx^\la\otimes\dr_i
\]
with respect to some connection (\ref{5})
on $Y\to X$. It is the fact why the gauge principle and the variation
principle involve connections on fibred manifolds in order
to describe field systems. In case of the standard gauge theory, they are
principal connections treated the mediators of interaction of fields.

Note that several equivalent definitions of connections on
fibred manifolds are utilized. We follow the general notion of
connections as sections
of jet bundles, without appealing to transformation groups. This
general approach is suitable to formulate the classical concept of
principal connections as sections of the jet bundle $J^1P\to P$ which
are equivariant under the canonical action of the structure group $G$
of $P$ on $P$ on the right. Then, we get description of principal
connections as sections of the bundle (\ref{3}).

\bigskip
{\bf II.}
\bigskip

In comparison with case of internal symmetries, the gauge
gravitation theory meets two objects. These are the fermion matter and
the geometric arena. They might arise owing to {\it sui generis} primary
phase transition which had separated prematter and pregeometry. One
can think of the well-known possibility of describing a space-time in
spinor coordinates as being the relic of that phase transition.

Here,
we are not concerned with these quetions, and by a world manifold is meant
a 4-dimensional oriented manifold $X^4$ coordinatized in the standard manner.
By world geometry is called differential geometry of the tangent bundle
$TX$ and the cotangent bundle $T^*X$ of $X^4$. The structure group of
these bundles is the general linear group
\[
GL_4=GL(4,\Bbb R).
\]
The associated principal bundle is the bundle $LX\to X$ of linear frames
in tangent spaces to $X^4$.

Note that, on the physical level, a basis of the tangent space $T_x$ to
$X^4$ at $x\in X^4$ is usually interpreted as the local reference frame
at a point.
However, realization of such a reference frame by physical devices
remains under discussion. A family $\{z_\xi\}$ of local sections of the
principal bundle $LX$ sets up an atlas of $LX$ (and so the associated atlases
of $TX$ and $T^*X$) which we, following the gauge theory tradition, treat
{\it sui generis} the world reference frame. In gauge gravitation
theory in comparison with case of internal symmetries, there exists the
special subclass of holonomic atlases $\{\psi^T_\xi\}$ whose
trivialization morphisms  are the tangent morphisms
\[
\psi^T_\xi=T\chi_\xi
\]
to trivialization morphisms $\chi_\xi$ of coordinate atlases of $X^4$. The
associated bases of $TX$ and $T^*X$ are the holonomic bases
$\{\dr_\la\}$ and $\{dx^\la\}$ respectively. Note that Einstein's
General Relativity was formulated just with respect to holonomic
reference frames, so that many mixed up reference frames and coordinate
systems. The mathematical reason of separating holonomic reference frames
consists in the fact that the jet manifold formalism is phrased only in
terms of holonomic atlases. However, holonomic atlases are not
compatible generally with spinor bundles whose sections describe the fermion
fields.

Different spinor models of the fermion matter have been suggested. But all
observable fermion particles are the Dirac fermions. There are several
ways in order to introduce Dirac fermion fields. We do this as follows.

Given
a Minkowski space $M$ with the Minkowski metric $\eta$, let
$\Bbb C_{1,3}$ be the complex Clifford algebra generated by elements
of $M$.
A spinor space $V$ is defined to be a minimal left ideal of $\Bbb C_{1,3}$ on
which this algebra acts on the left. We have the representation
\begin{equation}
\g: M\otimes V \to V \label{521}
\end{equation}
of elements of the Minkowski space $M$ by
Dirac's matrices $\g$ on $V$.
Let us consider the transformations preserving the representation (\ref{521}).
These are pairs $(l,l_s)$ of Lorentz transformations $l$ of  the Minkowski
space $M$ and invertible elements $l_s$ of $\Bbb C_{1,3}$ such that
\[\g (lM\otimes l_sV)=l_s\g (M\otimes V).\]
Elements $l_s$  form the Clifford group whose action on $M$
however is not effective, therefore we restrict ourselves to its spinor
subgroup $L_s =SL(2,\Bbb C)$.

Let us consider a bundle of complex Clifford algebras $\Bbb C_{3,1}$
over $X^4$. Its subbundles are both a spinor bundle $S_M\to X^4$ and the
bundle $Y_M\to X^4$ of Minkowski spaces of generating elements of
$\Bbb C_{3,1}$.
To describe Dirac fermion fields on a world manifold, one must
require that $Y_M$ is isomorphic to the cotangent bundle $T^*X$
of a world manifold $X^4$. It takes place if the structure group of
$LX$ is reducible to the Lorentz group
$L=SO(3,1)$ and $LX$
contains a reduced $L$ subbundle $L^hX$ such that
\[Y_M=(L^hX\times M)/L.\]
In this case, the spinor bundle $S_M$ is associated with the $L_s$-lift
$P_h$ of $L^hX$:
\begin{equation}
S_M=S_h=(P_h\times V)/L_s.\label{510}
\end{equation}

In accordance with the well-known theorem, there is the 1:1 correspondence
between the reduced subbubdles $L^hX$ of $LX$ and
the tetrad gravitational fields $h$ identified with global sections
of the quotient bundle
\[
\Si:= LX/L\to X^4.
\]
This bundle is the 2-fold cover of the bundle of pseudo-Riemannian
bilinear forms in cotangent spaces to $X^4$. Global sections of the
latter are pseudo-Riemannian metrics $g$ on $X^4$. Thus, existence of a
gravitational field is necessary condition in order that Dirac fermion
fields live on a world manifold.

Note that a world manifold $X^4$ must satisfy
the well-known global topological conditions in order that gravitational
fields, space-time structure and spinor structure can exist. To
summarize these conditions, one may assume that $X^4$ is not compact and
the linear frame bundle $LX$ over $X^4$ is trivial.

Given a tetrad field $h$, let $\Psi^h$ be an atlas of
$LX$ such that the corresponding local sections $z_\xi^h$ of $LX$
take their values into the reduced subbundle $L^hX$. This atlas has
$L$-valued transition functions. It fails to be holonomic in general.
Moreover, relative to $\Psi^h$, the pseudo-Riemannian metric $g$
corresponding to $h$ comes to the Minkowski metric and $h$ takes its
values in the center $\si_0$ of the quotient space $GL_4/L$.
With respect to an atlas $\Psi^h$ and a
holonomic atlas $\Psi^T=\{\psi_\xi^T\}$ of $LX$, the tetrad field $h$
can be represented by a family of $GL_4$-valued tetrad functions
\be
&& h_\xi=\psi^T_\xi\circ z^h_\xi,\\
&&dx^\la= h^\la_a(x)h^a,
\ee
which carry atlas (gauge) transformations between fibre bases
$\{dx^\la\}$ and $\{h^a\}$ of $T^*X$ associated with $\Psi^T$ and $\Psi^h$
respectively. The well-known relation
\begin{equation}
g^{\mu\nu}=h^\mu_ah^\nu_b\eta^{ab} \label{9}
\end{equation}
takes place.

It is the feature only of a tetrad
gravitational field that it itself determines reference frames.

Given a tetrad field $h$, one can define the representation
\begin{equation}
\g_h: T^*X\otimes S_h=(P_h\times (M\otimes V))/L_s\to (P_h\times
\g(M\otimes V))/L_s=S_h, \label{L4}
\end{equation}
\[
\wh dx^\la=\g_h(dx^\la)=h^\la_a(x)\g^a,
\]
of cotangent vectors to a world manifold $X^4$ by Dirac's $\g$-matrices
on elements of the spinor bundle $S_h$.

Let $A_h$ be a connection on $S_h$ associatede with a principal
connection on $L^hX$ and $D$
the corresponding covariant differential. Given the
representation (\ref{L4}), one can construct the Dirac operator
\[
\cD_h=\g_h\circ D: J^1S_h\to T^*X\op\otimes_{S_h}VS_h\to VS_h
\]
on $S_h$. Then, we can say that sections of
the spinor bundle $S_h$ describe Dirac fermion fields in the presence of
the tetrad gravitational field $h$ and the gauge potential $A_h$.

Thus, the geometry of the gauge gravitation theory is the metric-affine
geometry characterized by the pair $(h,A_h)$ of a tetrad
field $h$ and a reduced Lorentz connection $A_h$ treated as the gauge
gravitational potential. Note that the metric-affine geometry is an
attribute of all gauge approaches to gravitation theory \cite{heh,mil}.
Moreover, it is also the Klein-Chern geometry of Lorentz invariants.

\bigskip
{\bf III.}
\bigskip

If to forget fermion fields for a moment, one can derive the gauge
gravitation theory from the equivalence principle reformulated in
geometric terms \cite{iva,3sar}.

In Einstein's General Relativity, the equivalence principle is called to
provide transition to Special Relativity with respect to some reference
frames. In the spirit of F.Klein's Erlanger program, the Minkowski space
geometry can be characterized as geometry of Lorentz invariants. The
geometric equivalence principle then postulates that there exist reference
frames with respect to wich Lorentz invariants can be defined everywhere
on a world manifold $X^4$. This principle has
the adequate mathematical formulation in terms of fibre bundles.
It requires that the linear frame bundle
$LX$ is reducible to a Lorentz subbundle $L^hX$ whose
structure group is $L$. They are atlases of $L^hX$ with respect to which
Lorentz invariants can be defined, and the pseudo-Riemannian metric $g$
corresponding to $h$ exemplifies such a Lorentz invariant.

Thereby, the geometric equivalence principle provides a world manifold
with the so-called $L$-structure \cite{sul}. From the physical point of
view, it singles out the Lorentz group as the exact symmetry subgroup
of world symmetries broken spontaneously \cite{iva}. The associated
classical Higgs field is a tetrad gravitational field.

Note that the geometric equivalence principle sets au also a space-time
structure on a world manifold $X^4$. In virtue of the well-known theorems,
if the structure group of $LX$ is reducible to the structure Lorentz group,
the latter, in turn, is reducible to its maximal compact subgroup $SO(3)$.
Moreover, we have the commutative diagram
\begin{equation}
\begin{array}{ccc}
 GL_4 &  \longrightarrow & SO(4)  \\
\put(0,10){\vector(0,-1){20}} & & \put(0,10){\vector(0,-1){20}} \\
 L & \longrightarrow & SO(3)
\end{array} \label{8}
\end{equation}
of structure groups of $LX$.
It follows that, for every reduced subbundle
$L^hX$, there exist a reduced subbundle $FX$ of $LX$ with the
structure group $SO(3)$ and the corresponding (3+1) space-time decomposition
\[
TX = FX\oplus T^0X
\]
of the tangent bundle of $X^4$ into a 3-dimensional
spatial distribution $FX$ and its time-like orthocomplement $T^0X$.

In other words, we have also the Klein-Chern geometry of spatial
invariants on a world manifold. One can always choose an atlas of $LX$
whose transition functions are $SO(3)$-valued, and spatial invariants
are exemplified by the global field of local time directions $T^0X$.

Recall that there is the 1:1 correspondence
\[FX\rfloor \Omega = 0\]
between the nonvanishing 1-forms $\Omega$ on a manifold $X$ and
the 1-codimensional distributions on $X$.

Then, we get the following modification of the well-known theorem
\cite{3sar,1sard}.
\begin{itemize}
\item For every gravitational field $g$ on a world  manifold
$X^4$, there exists an associated pair $(FX,g^R)$ of a
space-time distribution $FX$ generated by a tetrad 1-form
\[
h^0=h^0_\mu dx^\mu
\]
and a Riemannian metric $g^R$, so that
\begin{equation}
g^R=2h^0\otimes h^0-g. \label{2.1}
\end{equation}
Conversely,
given a Riemannian metric $g^R$,  every  oriented  smooth  3-dimensional
distribution $FX$ with a generating form $\Omega$ is a space-time
distribution compatible with the gravitational field $g$ given by
expression (\ref{2.1}) where
\[
h^0=\frac{\Omega}{|\Omega|},\qquad |\Omega|^2=g^R(\Omega,\Omega) =
g(\Omega,\Omega).
\]
\end{itemize}

The triple $(g,FX,g^R)$ (\ref{2.1})
sets up uniquely a space-time structure on a
world manifold. In particular, if the generating form of a space-time
distribution $FX$ is exact, we
have the causal space-time foliation of $X^4$ what corresponds exactly
to the stable causality by Hawking.

A Riemannian metric $g^R$ in the triple (\ref{2.1}) defines a
$g$-compatible distance function on a
world manifold $X^4$.  Such a function turns $X^4$ into a metric space whose
locally Euclidean topology is equivalent to the manifold topology on $X^4$.
Given a gravitational field $g$, the $g$-compatible Riemannian metrics and
the corresponding distance functions are different for different space-time
distributions $FX$. It follows that physical observers
associated  with different distributions perceive the same world manifold
as different Riemannian spaces.  The well-known relativistic changes of
sizes of moving bodies exemplify this phenomenon.

One often loses sight of the fact that a certain Riemannian
metric and, consequently, a metric topology can be associated with a
gravitational field \cite{3sar}. For instance,
one attempts to derive a
world topology directly from pseudo-Riemannian structure of a space-time.
These are path topology etc. \cite{haw}. If a  space-time  obeys  the
strong causality condition, such topologies coincide with the familiar
manifold topology of $X$. In general case, they however are rather
extraordinary.

\bigskip
{\bf IV.}
\bigskip

A glance on the diagramm (\ref{8}) shows that, in the gravitation
theory, we have a collection of spontaneous symmetry breakings:
\begin{itemize}
\item $GL_4\to L$ where the corresponding Higgs field is a
gravitational field;
\item $L\to SO(3)$ where the Higgs-like field is represented by the
tetrad 1-form $h^0$ as a global section of the quotient bundle
\[
L^hX/SO(3)\to X^4;
\]
\item $GL_4\to SO(4)$ where the Higgs-like field is the Riemannian metric
>from expression (\ref{2.1}).
\end{itemize}

Spontaneous symmetry breaking is quantum phenomenon modelled by a
Higgs field. In the algebraic quantum field theory, Higgs fields
characterize nonequivalent Gaussian states of algebras of quantum fields.
They are {\it sui generis} fictitious fields describing
collective phenomena.
In the gravitation theory, spontaneous symmetry breaking displays on the
classical level and the feature of a gravitational field is that it is
a dynamic Higgs field. Indeed, the splitting (\ref{9}) of the metric
field looks like the standard decomposition of a Higgs field where the
Minkowski metric $\eta$ and the tetrad functions play the role of the
$L$-stable vacuum Higgs field and the Goldstone fields respectively.
However, in contrast with the internal symmetry case, the Goldstone
components of a gravitational field can not be removed by gauge
transformations because the reference frames $\Psi^h$ fail to be
holonomic in general and, roughly speaking, the associated basis elements
$h_a=h^\mu_a\dr_\mu$ contain tetrad functions.

>For the first time, the conception of a graviton as a Goldstone particle
corresponding to violation of Lorentz symmetries in a curved space-time
had been advanced in mid 60s by Heisenberg and Ivanenko in discussion on
cosmological and vacuum asymmetries. This idea was revived in connection
with constructing the induced representations of the group $GL_4$
and then in the framework of the approach to
gravitation theory as a nonlinear $\si$-model \cite{heh,nee,per}. In geometric
terms, the fact that a pseudo-Riemannian metric is similar to a Higgs field
has been pointed out by A.Trautman and by us \cite{iva}.

The Higgs character of classical gravity is founded on the fact that,
for different tetrad fields $h$ and $h'$, Dirac fermion fields are
described by sections of spinor bundles associated
with different reduced $L$-principal subbundles
of $LX$ and so, the representations $\gamma_h$ and $\gamma_{h'}$
(\ref{L4}) are not equivalent \cite{3sar}.
It follows that e
Dirac fermion field must be regarded only in a pair with a certain
tetrad gravitational field $h$. These pairs constitute the so-called
fermion-gravitation complex \cite{nee}. They can not be represented by
sections of any product $S\times\Si$ where $S\to X^4$ is some standard
spinor bundle. At the same time, there is the 1:1 correspondence between
these pairs and the sections of the composite spinor bundle
\begin{equation}
S\to\Si\to X^4 \label{L1}
\end{equation}
where $S\to\Si$ is a spinor bundle associated with the $L$ principal
bundle $LX\to\Si$ \cite{3sar,sard10}. In particular, every spinor bundle
$S_h$ (\ref{510}) is isomorphic to restriction of $S$ to $h(X^4)\subset\Si$.

\bigskip
{\bf V.}
\bigskip

By a composite manifold is meant the composition
\begin{equation}
 Y\to \Si\to X \label{I1}
\end{equation}
where $Y\to\Si$ is a bundle denoted by $Y_\Si$ and $\Si\to X$ is a
fibred manifold.

In analytical mechanics, composite manifolds
\[
Y\to\Si\to\Bbb R
\]
characterize systems with variable parameters, e.g.
the classical Berry's oscillator \cite{6sar}.
In gauge theory, composite manifolds
\[
P\to \Si_K\to X
\]
describe spontaneous symmetry breaking \cite{2sar,sard}.

Application of composite manifolds to field theory is
founded on the following speculations. Given
a global section $h$ of $\Sigma$, the restriction $ Y_h$
of $Y_\Sigma$ to $h(X)$ is a fibred submanifold
of $Y\to X$. There is the 1:1 correspondence between
the global sections $s_h$ of $Y_h$ and the global sections of
the composite manifold (\ref{I1}) which cover $h$.
Therefore, one can say that sections $s_h$ of $Y_h$
describe fields in the presence of a background parameter
field $h$, whereas sections
of the composite manifold $Y$ describe all pairs $(s_h,h)$.
It is important when the bundles $Y_h$ and $Y_{h\neq h'}$ fail to be
equivalent in a sense. The configuration space of these pairs is the
first order jet manifold $J^1Y$ of the composite manifold $Y$.

The feature of the dynamics of field systems on composite manifolds
consists in the following \cite{sard,sard11}.

Let $Y$ be a composite manifold (\ref{I1})
provided with the fibred coordinates $( x^\la ,\si^m, y^i) $ where
$( x^\la ,\si^m)$ are fibred coordinates of $\Si$.
Every connection
\[
A_\Sigma=dx^\lambda\otimes(\dr_\lambda+\wt A^i_\lambda\dr_i)
+ d\sigma^m\otimes(\dr_m+A^i_m\dr_i)
\]
on $Y\to\Sigma$ yields splitting
\[
VY=VY_\Sigma\op\oplus_Y (Y\op\times_\Sigma V\Sigma)
\]
and, as a consequence,
the first order differential operator
\be
&&\wt D: J^1Y\to T^*X\op\otimes_Y VY_\Si, \\
&&\wt D= dx^\la\otimes(y^i_\la-\wt A^i_\la -A^i_m\si^m_\la)\dr_i,
\ee
on $Y$. Let $h$ be a global section
of $\Si$ and $Y_h$ the restriction of the bundle $Y_\Si$ to $h(X)$. The
restriction of $\wt D$ to $J^1Y_h\subset J^1Y$
comes to the familiar covariant differential relative to a certain
connection $A_h$ on $Y_h$.
Thus, it is $\wt D$ that
one may utilize in order to construct a Lagrangian density (\ref{20})
\[
L: J^1Y\op\to^{\wt D} T^*X\op\otimes_YVY_\Si\to\op\w^nT^*X
\]
for sections of a composite manifold. It should be noted that such a
Lagrangian density is never regular because of the constraint conditions
\[
A^i_m\dr^\mu_i\cL =\dr^\mu_m\cL.
\]

Recall that, if a Lagrangian density is degenerate, the corresponding
Euler-Lagrange equations are underdetermined and need supplementary
gauge-type conditions which remain elusive in general.
Therefore, to describe constraint field systems, we
utilize the multimomentum Hamiltonian formalism where
canonical momenta correspond to derivatives of fields with respect
to all world coordinates, not only the temporal one
\cite{car,gun,6sar,sard,bsar,lsar}. Note that application of
the conventional Hamiltonian formalism to field theory
fails to be successful. In the straightforward manner, it leads
to infinite-dimensional phase spaces.
In the framework of the multimomentum approach,
the phase space of fields is the Legendre bundle
\begin{equation}
\Pi=\op\w^n T^*X\op\otimes_Y TX\op\otimes_Y V^*Y \label{00}
\end{equation}
over $Y$. It is provided with the fibred coordinates $(x^\la ,y^i,p^\la_i)$.
Note that
every Lagrangian density  $L$ on $J^1Y$ determines the Legendre morphism
\be
&& \wh L: J^1Y\to \Pi,\\
&& (x^\m,y^i,p^\m_i)\circ\wh L=(x^\m,y^i,\dr^\mu_i\cL).
\ee

The Legendre bundle (\ref{00}) carries the multisymplectic form
\[
\Om =dp^\la_i\w
dy^i\w\om\otimes\dr_\la.
\]
We  say that a connection
$\g$ on the fibred Legendre manifold $\Pi\to X$ is a Hamiltonian
connection if the  form  $\g\rfloor\Om$ is closed.
Then, a Hamiltonian form $H$ on $\Pi$ is defined to be an
exterior form such that
\begin{equation}
dH=\g\rfloor\Om \label{013}
\end{equation}
for some Hamiltonian connection $\g$. The key point consists in the
fact that every Hamiltonian form admits splitting
\begin{equation}
H =p^\la_idy^i\w\om_\la
-p^\la_i\G^i_\la\om
-\wt{\cH}_\G\om=p^\la_idy^i\w\om_\la-\cH\om,
\qquad \om_\la=\dr_\la\rfloor\om,\label{017}
\end{equation}
where $\G$ is a connection on the fibred manifold $Y$ and
$\wt{\cH}_\G\om$ is a horizontal density on $\Pi\to X$.
Given the  Hamiltonian form (\ref{017}), the equality
(\ref{013}) comes to the Hamilton equations
\ben
&&\dr_\la r^i =\dr^i_\la\cH, \nonumber\\
&& \dr_\la r^\la_i =-\dr_i\cH \label{3.11}
\een
for sections $r$ of the fibred Legendre manifold $\Pi\to X$.

We thus observe that the multimomentum Hamiltonian formalism
exemplifies the generalized Hamiltonian dynamics which is not merely a
time evolution directed by the Poisson bracket, but it is governed by
partial differential equations (\ref{3.11})
where temporal and spatial coordinates
enter on equal footing. Maintaining covariance has the principal
advantages of describing field theories, for any preliminary
space-time splitting shades the covariant picture of field constraints.
Contemporary field models are almost always the constraint ones.

We shall say that the Hamiltonian form $H$ (\ref{017}) is associated with a
Lagrangian density $L$ (\ref{6}) if $H$ satisfies the relations which
take the coordinate form
\[
\cH(y^j,p_j^\mu)=p_i^\la\dr^i_\la\cH -\cL(y^j,\dr^j_\mu\cH).
\]
If a Lagrangian density $L$ is regular, there exists the unique Hamiltonian
form $H$ such that the first order Euler-Lagrange equations and the
Hamilton equations are equivalent, otherwise in general case.
One must consider a
family of different  Hamiltonian forms $H$ associated with the same
degenerate Lagrangian
density $L$ in order to exaust solutions of the Euler-Lagrange
equations.
Lagrangian densities of field models are almost always quadratic and
affine in derivative coordinates $y^i_\mu$.
In this case,
given an associated Hamiltonian form $H$, every solution of the
corresponding Hamilton equations which
lives on $\wh L(J^1Y)\subset\Pi$ yields
a solution of the Euler-Lagrange equations. Conversely,
for any solution of the Euler-Lagrange equations, there
exists the corresponding solution of the Hamilton equations for some
associated Hamiltonian form. Obviously, it lives on  $\wh L(J^1Y)$ which
makes the sense of the Lagrangian constraint space.

The feature of Hamiltonian systems on composite manifolds (\ref{I1})
lies in the fact that the Lagrangian constraint space is
\begin{equation}
p^\la_m+A^i_mp^\la_i=0 \label{502}
\end{equation}
\cite{sard,bsar,sard11}.
Moreover, if $h$ is a global section of $\Si\to X$, the submanifold
$\Pi_h$ of $\Pi$ given by the coordinate relations
\[
\si^m=h^m(x), \qquad p^\la_m+A^i_mp^\la_i=0
\]
is isomorphic to the Legendre bundle over the restriction $Y_h$ of
$Y_\Si$ to $h(X)$. The Legendre bundle $\Pi_h$ is the phase space of
fields in the presence of the background parameter field $h$.

\bigskip
{\bf VI.}
\bigskip

In gravitation theory, we have the composite manifold
\[
LX\to\Si\to X^4
\]
and the associated composite spinor bundle (\ref{L1}). Roughly
speaking, values of tetrad gravitational fields play the role of
coordinate parameters, besides the familiar world coordinates.

In the multimomentum Hamiltonian gravitation theory,
the constraint condition (\ref{502})
takes the form
\begin{equation}
p^{c\la}_\mu+\frac18\eta^{cb}\si^a_\mu(y^B[\g_a,\g_b]^A{}_B
p^\la_A+p^{A\la}_+[\g_a,\g_b]^{+B}{}_Ay^+_B)=0 \label{M2}
\end{equation}
where $(\si^\mu_c,y^A)$ are tetrad and spinor coordinates of the
composite spinor bundle (\ref{L1}), and $p^{c\la}_\mu$ and
$p^\la_A$ are the corresponding momenta \cite{sard10,sard11}.
The condition (\ref{M2}) replaces the standard gravitational constraints
\begin{equation}
p^{c\la}_\mu=0. \label{M1}
\end{equation}
The crucial point is that, when restricted
to the constraint space (\ref{M1}), the Hamilton equations (\ref{3.11})
of gravitation theory
come to the familiar gravitational equations
\[
G^a_\mu +T^a_\mu=0
\]
where $T$ denotes the energy-momentum tensor of fermion fields,
otherwise on the modified constraint space (\ref{M2}). In the latter case,
we have the modified gravitational equations
of the total system of fermion fields and gravity:
\[
D_\la p^{a\la}_\mu=G^a_\mu +T^a_\mu
\]
where $D_\la$ denotes the covariant derivative with respect to the
Levi-Civita connection which acts on the indices $^a_\mu$.

\bigskip
{\bf VII.}
\bigskip

Since, for different tetrad fields $h$ and $h'$,
the representations $\gamma_h$ and $\gamma_{h'}$
(\ref{L4}) are not equivalent, even weak gravitational fields, unlike
matter fields and gauge potentials, fail to form an affine space
modelled on a linear space of deviations of some background field.
They thereby do not satisfy the superposition principle and can not be
quantized by usual methods, for in accordance with the algebraic
quantum field theory quantized fields must constitute a linear space.
This is the common feature of Higgs fields. In algebraic quantum field
theory, different Higgs fields correspond to nonequivalent Gaussian
states of a quantum field algebra. Quantized deviations of a Higgs
field can not change a state of this algebra and so, they fail to
generate a new Higgs field.

At the same time, one can examine
superposable deviations $\si$ of a tetrad gravitational field $h$ such
that $h+\si$ is not a tetrad gravitational field \cite{sard90,3sar}.
In the coordinate form, such deviations read
\ben
&&\wt h^\mu_a=H^b{}_ah^\mu_b = (\dl^b_a+\si^b{}_a)h^\mu_b=
H^\mu{}_\nu h^\nu_a =(\dl^\mu_\nu+\si^\mu{}_\nu)h^\nu_a=h^\mu_a+
\si^\mu{}_a, \label{3.5}\\
&&\wt h^a_\mu=g_{\mu\nu}\eta^{ab}\wt h^\nu_b=H_\mu{}^\al h^a_\al,
\nonumber\\
&&\wt h^\mu_a\wt h^a_\nu\neq \dl^\mu_\nu,\qquad \wt h^\mu_a\wt
h^b_\mu\neq \dl^b_a.\nonumber
\een
Note that the similar factors have been investigated by R.Percacci
\cite{per2}.

In bundle terms, we can describe the deviations (\ref{3.5}) as
the special morphism $\Phi_1$ of the cotangent bundle \cite{sard90,3sar}.
Given a gravitational field $h$ and the corresponding representation
morphism $\g_h$ (\ref{L4}), the morphism $\Phi_1$ yields another
$\g$-matrix representation
\be
&&\wt\g_h=\g_h\circ\Phi_1,\\
&&\wt\g_h(h^a)=H^a{}_b\g_h(h^b)=H^a{}_b\g^b,
\ee
of cotangent vectors, but on the same spinor bundle $S_h$. Therefore,
deviations (\ref{3.5}) and their superposition $\si + \si'$ can be defined.

Let us note that, to construct a Lagrangian density of deviations
$\epsilon$ of a
gravitational field, one usually utilize a familiar Lagrangian
density of a gravitationsl field $h' = h + \epsilon$
where $h$ is treated as a background field. In case of the
deviations (\ref{3.5}), one can not follow this method, for quantities
$\wt h$ fail to be true tetrad fields. To overcome this difficulty, we
use the fact that
the morphisms $\Phi_1$ appears also in the dislocation
gauge theory of the translation group.  We therefore may
apply the Lagrangian densities of this theory in order to describe
deviations $\si$ (\ref{3.5}).

Let the tangent bundle $TX$ be provided
with the canonical structure of the affine tangent bundle. It is
coordinatized by $(x^\mu, u^\la)$ where $u^\al\neq \dot x^\al$ are the
affine coordinates.

Every affine connection $A$ on $TX$ is brought into the sum
\begin{equation}
A=\G+\si \label{4.4}
\end{equation}
of a linear connection $\G$ and a soldering form
\[
\si=\si^\la{}_\mu(x)\dr_\la\otimes dx^\mu
\]
which plays the role of a gauge translation potential.

In the conventional gauge theory of the affine  group, one
faces the problem of physical interpretation of both gauge
translation potentials and sections $u(x)$ of the affine tangent
bundle $TX$. In field theory, no fields possess the transformation
law
\[u(x)\to u(x) + a\]
under the Poincar\'e translations.

At the same time, one observes such fields in the gauge theory
of dislocations \cite{kad} which is based on the fact that, in the presence
of dislocations, displacement vectors $u^k,\, k = 1,2,3,$
of small deformations are  determined  only  with  accuracy  to  gauge
translations
\[
u^k\to u^k+a^k(x).
\]
In this theory, gauge translation potentials $\si^k{}_i$ describe
the plastic distortion, the covariant derivatives
\[
D_iu^k =\dr_i u^k - \si^k{}_i
\]
consist with the elastic distortion, and the
strength
\[\cF^k{}_{ij}=\dr_i \si^k{}_j -\dr_j \si^k{}_i\]
the dislocation density.
Equations of the dislocation theory
are derived from the gauge invariant Lagrangian density
\begin{equation}
\cL=\mu D_iu^kD^iu_k+\frac{\la}{2} D_iu^iD_mu^m-\epsilon \cF^k{}_{ij}
\cF_k{}^{ij} \label{4.7}
\end{equation}
where $\mu$ and $\lambda$ are the Lame coefficients of isotropic media.
These equations however are not
independent of each other since a displacement field
$u^k(x)$ can be removed by gauge translations and, thereby, it fails to be a
dynamic variable.

In the spirit of the gauge dislocation theory, we have suggested
that gauge  potentials of the Poincar\'e  translations  may  describe  new
geometric structure ({\it sui generis} dislocations) of a world
manifold \cite{sard90,3sar}.

Let the tangent bundle $TX$ be provided with an affine
connection (\ref{4.4}).
By dislocation morphism of a world manifold
$X$ is meant the special bundle isomorphism of $TX$ over $X$ which
takes the coordinate form
\begin{equation}
\rho: \tau^\mu\frac\dr{\dr x^\mu}\to \frac\dr{\dr x^\mu}+
(\G^\al{}_{\bt\mu}u^\bt+\si^\al{}_\mu)\frac\dr{\dr
u^\al}\to\tau^\mu (\dl^\al_{\mu}+\si^\al{}_\mu)
\frac\dr{\dr x^\al}=\tau^\mu H^\al{}_\mu\frac\dr{\dr x^\al},\label{4.8}
\end{equation}
where
\begin{equation}
\si^\al{}_\mu=D_\mu u^\al|_{u=0}=(\dr_\mu u^\al+\G^\al{}_{\bt\mu}
u^\bt+\si^\al{}_\mu)|_{u=0} \label{11}
\end{equation}
is the covariant derivatives of a displacement field $u$.

Let $Y$ be a bundle over $X$ and $J^1Y$ the jet manifold of $Y$.
The deformation morphism~(\ref{4.8}) has the jet prolongation
\[
j^1\rho: y^i_\la\to H^\al{}_\la(x)y^i_\al
\]
over $Y$. Then, given a Lagrangian density
$L$ of fields on a world manifold, one can think of the composition
\[
\wt L=L\circ \rho
\]
as being the corresponding Lagrangiam density of fields on the
dislocated manifold. If the above-mentioned morphism $\Phi_1$
consists with the dual to
the morphism (\ref{4.8}), one can apply
these Lagrangian densities in order to describe fields in the presence of
the deviations (\ref{3.5}). Moreover, we may assume that
the deviations $\si$ (\ref{3.5}) have the dislocation nature (\ref{11}).

Note that a Lagrangian density $L_{(\si)}$ of translation gauge potentials
$\si^\la{}_\mu$ cannot be built in the standard Yang-Mills form since the
Lie algebra of the affine group does not admit an invariant nondegenerate
bilinear form. To construct $L_{(\si)}$, one can utilize the torsion
\[
\cF^\al{}_{\nu\mu}=D_\nu\si^\al{}_\mu  -D_\mu\si^\al{}_\nu
\]
of the linear connection $\G$ with respect to the soldering form $\si$.
The general form of a Lagrangian
density $L_{(\si)}$ is given by the expression
\be
&&\cL_{(\si)}=\frac12[a_1\cF^\mu{}_{\nu\mu} \cF_\al{}^{\nu\al}+
a_2\cF_{\mu\nu\si}\cF^{\mu\nu\si}+a_3\cF_{\mu\nu\si}\cF^{\nu\mu\si}\\
&&\qquad+a_4\epsilon^{\mu\nu\si\g}\cF^\epsilon{}_{\mu\epsilon}
\cF_{\g\nu\si}-\mu\si^\mu{}_\nu\si^\nu{}_\mu+
\la\si^\mu{}_\mu \si^\nu{}_\nu]\sqrt{-g}.
\ee

This Lagrangian density differs from the familiar gravitational
Lagrangian densities. In particular, it contains the
mass-like term originated from the Lagrangian
density (\ref{4.7}) for displacement fields $u$ under the gauge condition
$u=0$. Solutions of the corresponding field equations show that fields $\si$
make contribution to the standard gravitational effects. In particular,
they lead to the "Yukawa type" modification of Newton's gravitational
potential.


\begin{thebibliography}{ederf}

\bibitem{car} J. Cari\~nena, M. Crampin and L. Ibort, {\it Diff.
Geom. Appl.}, {\bf 1}, 345 (1991).

\bibitem{gia} G. Giachetta and L. Mangiarotti, {\it Int.
J. Theor. Phys.}, {\bf 29}, 789 (1990).

\bibitem{got} M. Gotay, in
{\it Mechanics, Analysis and Geometry: 200 Years after Lagrange}, ed.
M.Francaviglia (Elseiver Science Publishers B.V., 1991) p. 203.

\bibitem{gun} C. G\"unther, {\it J. Diff. Geom.}, {\bf 25}, 23 (1987).

\bibitem{haw} S. Hawking and G. Ellis, {\it The Large Scale Structure of
a Space-Time} (Cambrifge Univ. Press, Cambridge, 1973).

\bibitem{heh} F. Hehl, J. McCrea, E. Mielke and Y. Ne'eman: Metric-affine
gauge theory of gravity, {\it Physics Reports} (1995) (to appear).

\bibitem{iva} D. Ivanenko and G. Sardanashvily, {\it Physics Reports},
{\bf 94}, 1 (1983).

\bibitem{kad} A. Kadic and D. Edelen, {\it A Gauge Theory of Dislocations
and Disclinations} (Springer, New York, 1983).

\bibitem{kol} I. Kola\v{r}, P.W. Michor, J. Slov\'ak,
{\it Natural operations in differential geometry}, (Springer-Verlag, Berlin
etc. 1993).

\bibitem{kup} B. Kupershmidt, {\it Geometry of Jet Bundles and the
Strucuture of Lagrangian and Hamiltonian Formalisms}, Lect. Notes
in Math., {\bf 775}, 162 (1980).

\bibitem{man} L. Mangiarotti and M. Modugno, in {\it Geometry and
Physics}, ed. M.Modugno (Pitagora Editrice, Bologna, 1982) p. 135.

\bibitem{mil} E. Mielke: {\it Geometrodynamics of Gauge Fields - On the
Geometry of Yang-Mills and Gravitational Gauge Theories}
(Akademic-Verlag, Berlin, 1987).

\bibitem{nee} J. Ne'eman and Dj. \v Sija\v cki, {\it Annals of Phsics},
{\bf 120}, 292 (1979).

\bibitem{per} R. Percacci, {\it Geometry of Nonlinear Field Theories}
(World Scientific, Singapore, 1986).

\bibitem{per2} R. Percacci, {\it Nuclear Physics}, {\bf B353}, 271 (1991).

\bibitem{sard90} G.Sardanashvily, {\it Acta
    Physica Polonica}, {\bf B21}, 583 (1990).

\bibitem{2sar} G. Sardanashvily, {\it
J. Math. Phys.}, {\bf 33},  1546 (1992).

\bibitem{3sar} G. Sardanashvily  and O. Zakharov,  {\it Gauge
Gravitation Theory}  (World Scientific, Singapore, 1992).

\bibitem{6sar}
G. Sardanashvily and O. Zakharov, {\it Diff. Geom.
Appl.}, {\bf 3}, 245 (1993).

\bibitem{sard} G.Sardanashvily, {\it Gauge Theory in Jet Manifolds}
(Hadronic Press, Palm Harbor, 1993)

\bibitem{bsar} G. Sardanashvily, {\it Generalized Hamiltonian Formalism
for Field Theory. Constraint Systems} (World Scientific, Singapore, 1994).

\bibitem{lsar} G. Sardanashvily, Multimomentum Hamiltonian Formalism
in Field Theory,  E-print: hp-th/9403172, 9405040.

\bibitem{1sard} G. Sardanashvily, Gravitation Singularities of the
Caustic Type, E-print: gr-qc/9404024.

\bibitem{sard10} G. Sardanashvily, Gravity as a Higgs Field,
E-print: gr-qc/9405013, 9407032.

\bibitem{sard11} G.Sardanashvily, Hamiltonian field systems on
composite manifolds, E-print: hep-th/9409159.

\bibitem{sul} R. Sulanke and P. Wintgen, {\it Ddifferetialgeometrie
und Faserb\"undel} ( Veb Deutsher Verlag der Wissenschaften, Berlin,
1972).

\bibitem{sau} D. Saunders, {\it The Geometry of Jet Bundles},
(Cambridge Univ. Press, Cambridge, 1989).

\end{thebibliography}
 \end{document}